\documentclass[conference]{IEEEtran}
\IEEEoverridecommandlockouts

\usepackage{cite}
\usepackage{amsmath,amssymb,amsfonts}
\usepackage[hidelinks,urlcolor=blue]{hyperref} 
\usepackage{graphicx}
\usepackage{textcomp}
\usepackage{xcolor}
\usepackage{gensymb}
\usepackage{algorithm}
\usepackage{algorithmic}
\graphicspath{{figures/}}
\def\BibTeX{{\rm B\kern-.05em{\sc i\kern-.025em b}\kern-.08em
    T\kern-.1667em\lower.7ex\hbox{E}\kern-.125emX}}

\begin{document}

\title{Direct Time-of-Flight Measurement Accuracy Improvement With Perimeter-Gated SPADs

\thanks{This material is based on work supported by the National Science Foundation
under Grant No. 2442346. Any opinions, findings, and conclusions or recommendations expressed in this material are those of the authors and do not necessarily reflect the views of the National Science Foundation.}

}

\author{\IEEEauthorblockN{Md Sakibur Sajal, Hunter Guthrie, Zexi Liu and Marc Dandin}
{Department of Electrical and Computer Engineering},\\
{Carnegie Mellon University,}
{Pittsburgh, Pennsylvania, 15213, USA}\\
\\
{email: mdandin@andrew.cmu.edu}   }


\maketitle

\begin{abstract}
Direct time of flight (dToF) measurements are susceptible to errors because of system-level and circuit-level timing jitters. In addition, device-level uncertainty stemming from the dark noise of single-photon avalanche diode (SPAD) contributes to the aggregated error. We demonstrate that perimeter gating can help reduce the device-level detection inaccuracy for SPAD devices by reducing the dark noise probability. Specifically, in this work, we developed a general framework to accurately estimate the dToF jitters stemming from different source levels and analyzed a counter-based time to digital converter (TDC) circuit that are commonly used in such systems. We have also measured dToFs using a perimeter-gated SPAD (pg-SPAD) detector fabricated in a $\mathbf{0.35~\mu m}$ standard CMOS process. Experimental results show that pg-SPADs can improve measurement accuracy in both free-running and time-gated operations.
\end{abstract}


\section{Introduction}

Direct time-of-flights (dToFs) are measured using a \textit{stopwatch} mechanism that starts and stops the time measurements based on the emission of a laser pulse and the subsequent detection of its reflection, respectively~\cite{dtof1,dtof2,dtof3,dtof4,dtofintro1,dtofintro2,dtofintro3}. Time to digital converters (TDCs)~\cite{tdc1,tdc2,tdc3,tdc4} are used as a building block of these timer systems that precisely measure the pulse travel time. Hence, the accuracy of the dToF measurements primarily depends on the accuracy of the TDCs as well as the laser emitter and the photon detector's inherent timing uncertainty~\cite{dtofaccuracy1}. 

\begin{figure}[t]
    \centering
    \includegraphics[width=0.9\linewidth]{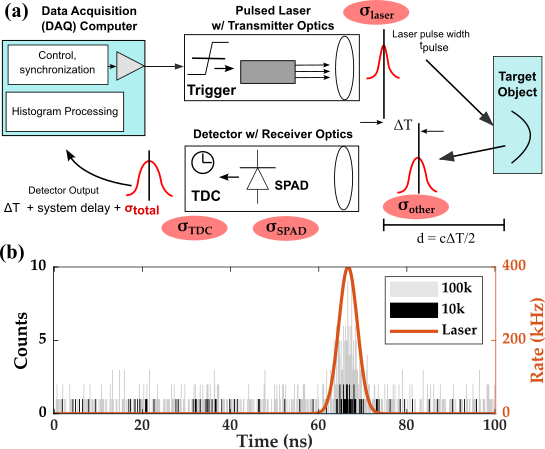}
    \vspace{-5pt}
    \caption{(a) Direct time-of-flight system architecture adapted from~\cite{dtof_model} and (b) simulated histograms using the first photon detection principle~\cite{first_photon}.}
    \vspace{-10pt}
    \label{fig:dtof_model}
\end{figure}

Figure~\ref{fig:dtof_model} (a) shows an adapted system model of a typical single-photon avalanche diode (SPAD) detector based dToF system~\cite{dtof_model}. A data acquisition (DAQ) computer sends a triggering signal to pulse a laser with a finite pulse width, $t_{pulse}$ toward a target object at a distance, $d$. The reflected photons are detected by a SPAD after a time delay of $\Delta T = \frac{2d}{c}$, where $c$ is the velocity of light in the surrounding medium. A TDC registers $\Delta T$ with a deterministic system delay in addition to the timing uncertainty $\sigma_{total}$ resulting from the jitters associated with the laser ($\sigma_{laser}$), the SPAD detector ($\sigma_{SPAD}$), the TDC ($\sigma_{TDC}$) and the environmental factors ($\sigma_{other}$) \textit{i.e.}, scattering from smoke, dust or fog. Since the uncertainty sources are independent, we can write~\cite{dtof_model} 
\begin{equation}
\label{eq:sigma}
    \sigma_{total} = \sqrt{\sigma^2_{laser}+\sigma^2_{other}+\sigma^2_{SPAD}+\sigma^2_{TDC}}.
\end{equation}

While there are several types of TDCs trading power, area and accuracy~\cite{6737636,6658231,6658237}, digital counter-based TDCs are one of the straightforward designs that can achieve sub-nanosecond timing jitters~\cite{cnttdc1,cnttdc2,cnttdc3}. As for the lasers, commercially available components can achieve jitters as low as tens of picoseconds. Ignoring the uncontrollable environmental factors~\cite{4292664}, the jitter of the SPAD detector can dominate $\sigma_{total}$ if the intrinsic dark noise probability overwhelms photon detection probability. In such cases, higher repetitions are needed to reduce the timing uncertainty.

\begin{figure*}
    \centering
    \includegraphics[width=\linewidth]{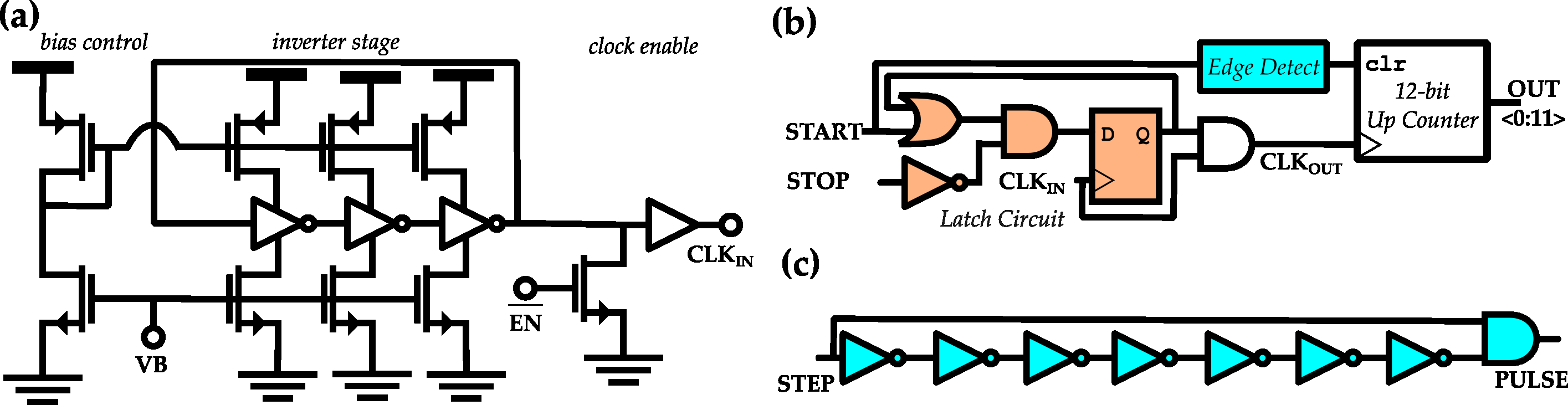}
    \vspace{-15pt}
    \caption{Architecture of a counter-based TDC with on-chip (a) clock generator, (b) 12-bit up counter, and (c) edge detector modules. The entire unit occupies $7000~\mu m^2$ in a $350~nm$ process. Clock frequency tunability is achieved by using a three-stage current starved inverter based ring oscillator.}
    \vspace{-15pt}
    \label{fig:tdc}
\end{figure*}

For instance, Figure~\ref{fig:dtof_model} (b) shows two simulated dToF histograms based on the first photon detection principle~\cite{first_photon} using a conventional SPAD device with a $50~kHz$ dark count rate (DCR). The laser pulse, with a width of $5~ns$ and a pulse rate of $400~kHz$, is being reflected from a target at $10~m$. We can clearly see a prominent histogram peak for a repetition of $100k$ samples. However, the peak drastically drops for a reduced number of repetitions, for instance, $10k$ samples which results in higher uncertainties in the dToF measurements unless the DCR is suppressed.

Although several dark noise reduction techniques are available for SPAD devices~\cite{Palubiak2011High-SpeedApplications, Finkelstein2006STI-BoundedTechnology, Richardson2011ScaleableTechnology}, perimeter gating is the only technology agnostic technique with the capability of dynamic noise modulation~\cite{Dandin2010, Nouri2012,  Dandin2012a, Dandin2016}. In this method, a polysilicon gate is laid over the perimeter junction of the diode, which is susceptible to premature edge breakdown~\cite{spadpuf}. By energizing this gate, the dark carrier generation at the perimeter junction can be actively reduced. This results in a decrease in the probability of dark noise~\cite{Sajal2022Perimeter-GatedProbability,Sajal2024TrueDiode}.

Several recent studies reported on the benefit of using perimeter-gated SPADs (pg-SPADs) as the photon detector in dToF systems~\cite{10405999,10115088}. However, these were simulation works based on fabricated pg-SPAD device parameters. This work bridges the gap by using a pg-SPAD detector with on-chip counter-based TDC to experimentally evaluate the effect of perimeter gate voltage on the accuracy of dToF measurements based on the first photon detection principle.

Specifically, we present a system-level, circuit-level, and device-level timing uncertainty analysis using a pg-SPAD array fabricated in a $0.35~\mu m$ standard CMOS process~\cite{Sajal2022Perimeter-GatedProbability}. An experimental setup, built based on the architecture shown in Fig.~\ref{fig:dtof_model} (a) was used with a pulse repetition of 1000 at $5~MHz$ to measure dToFs at different perimeter gate voltages. We observe a significant reduction in $\sigma_{SPAD}$ from the resulting histograms when compared to native SPAD devices. The benefit directly translated into improved signal fidelity when compared with that from a reduced number of pulse repetition.

\section{Architecture and Operation}
Figure~\ref{fig:tdc} shows the components of the counter-based TDC used in this work. A three-stage current starved inverter  based ring oscillator (see Fig.~\ref{fig:tdc} (a)) was designed to supply the clock signal, $\mathbf{CLK_{IN}}$. External bias control, $\mathbf{VB}$ can tune the clock frequency. A stable frequency range of $25~\sim150~MHz$ was observed for $\mathbf{VB} = 0.75~\sim~1.05~V$. An active-low enable signal can enable/disable the clock as needed.

Figure~\ref{fig:tdc} (b) shows the main TDC module. The \textbf{START} and the \textbf{STOP} signal pass through a latch circuit that freezes the clock signal, $\mathbf{CLK_{OUT}}$ to the counter when a photon or a dark carrier generated avalanche event is detected. At the beginning of the time measurement, an edge detect circuit (see Fig.~\ref{fig:tdc} (c)) clears the 12-bit up counter with a short pulse. After a settable active period, the counter output, $\mathbf{OUT<0:11>}$ is read out by the data acquisition system.

\begin{figure}[ht]
    \centering
    \includegraphics[width=\linewidth]{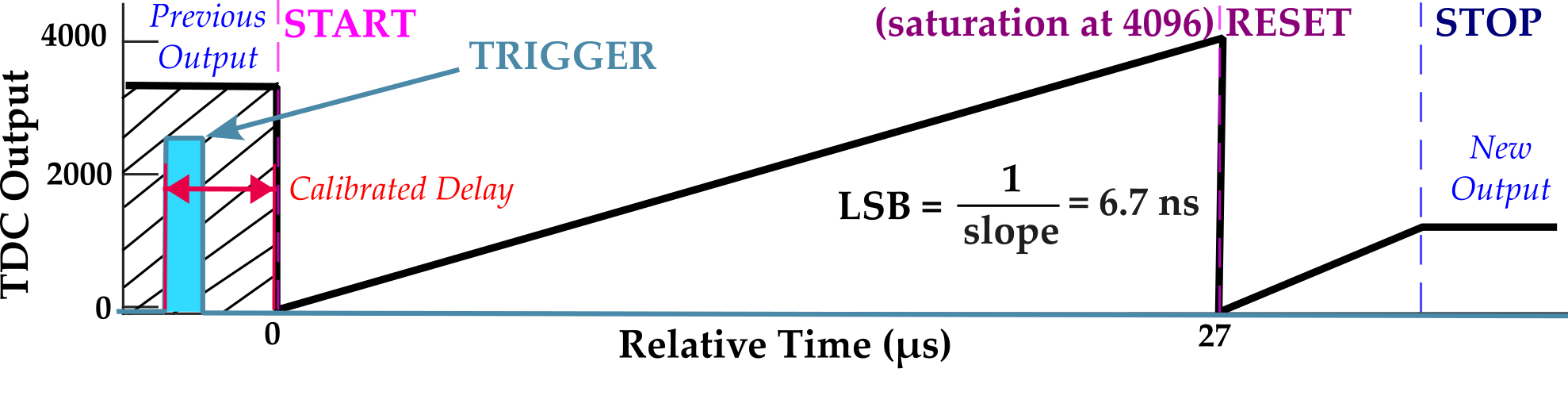}
    \vspace{-20pt}
    \caption{Timing diagram showing the TDC output during the default operation.}
    \vspace{-5pt}
    \label{fig:timing}
\end{figure}

Figure~\ref{fig:timing} shows the TDC output explaining its general operation. The output of the previous repetition is latched until the \textbf{START} signal for the new measurement is asserted, which resets the TDC and allows it to ramp up. For \textbf{VB} = 1.05 V, the clock frequency becomes $\sim150~MHz$, resulting in a TDC slope of $\frac{1}{6.7~ns}$ per code. Hence, it takes $27~\mu s$ (equivalent to $\sim4~km$ round trip by the laser pulse) to saturate the 12-bit TDC until it resets automatically and continues to ramp up. Upon an avalanche detection, caused by a photon or a dark carrier, the \textbf{STOP} signal is generated, which freezes the TDC to be read out by the DAQ. The dToF is calculated as the sum of complete ramp periods and the fraction of the final ramp based on the TDC output code. An independent \textbf{TRIGGER} signal triggers the laser before the \textbf{START} signal to balance out the signal delay from the DAQ to the laser's optical output.

\section{Materials and Methods}
We reproduced the setup shown in Fig.~\ref{fig:dtof_model} (a) using a data acquisition system (NI-PXIe 8821 controller with NI-PXIe 6544 digital waveform generator, National Instruments) and a 488 nm commercial nanosecond pulsed laser system with an adjustable pulse width of 6-39 ns (NLP49B, Thorlabs) to source the laser pulse. As for the detector, we have used a $\mathbf{64 \times64}$ pg-SPAD array previously reported in~\cite{Sajal2022Perimeter-GatedProbability}. Each $\mathbf{50\times50~\mu m^2}$ pixel with a $~6\%$ fill-factor contains a pg-SPAD and the in-pixel sensing circuitry. Output signal from the pg-SPADs are routed to the off-array TDCs through MUXs.

Deterministic delays and the system-level jitters were experimentally determined  to be calibrated out to properly investigate the effect of perimeter gate voltage on the measured dToFs. We used a high-speed digital oscilloscope (SDS1104X-E, Siglent) to collect the data for the system-level delays and jitters. Circuit-level delays and jitters within the pg-SPAD detector chip were estimated through post-layout simulation and experimental validation with an excess bias voltage of $\sim4~V$ at room temperature. 

\section{Experimental Observations and Discussions}
\subsection{System-level Uncertainty, $\sigma_{laser}$}
System-level timing uncertainties arise from the aggregated timing jitters of the DAQ module and the nanosecond pulsed laser system. Since we have used commercial components for this part, our reference values for system-level delays and jitters come from the manufacturer's datasheet (see Table~\ref{tab:jitter}). 

\begin{table}[]
    \centering
    \caption{System-level Delays and Jitters}
    \label{tab:jitter}
    \begin{tabular}{c|c|c|c} \hline
        Component & Manufacturer & Delay (ns)  & Jitter (ps)\\ \hline
        DAQ (PXIe-6547) & National Instruments & 0 (ref.) & $\pm 350$ \\
        Laser (NPL49B) & Thorlabs & $35 \pm5$ & $100~(max)$
    \end{tabular}

\end{table}

With the DAQ output as the time reference, we find that the channel-to-channel skew (\textit{i.e.}, the time difference between the \textbf{TRIGGER} and the \textbf{START} signals' rising edges) is $\pm 350~ps$. Incurring the cable delays from the DAQ to the laser trigger input port, these signals further experience a maximum RMS jitter of $100~ps$. We combine these two jitters as 
\begin{equation}
\label{eq:laser}
   \sigma_{laser} = \sqrt{(350)^2+(100)^2} \simeq 364~ps. 
\end{equation}

\begin{figure}
    \centering
    \includegraphics[width=\linewidth]{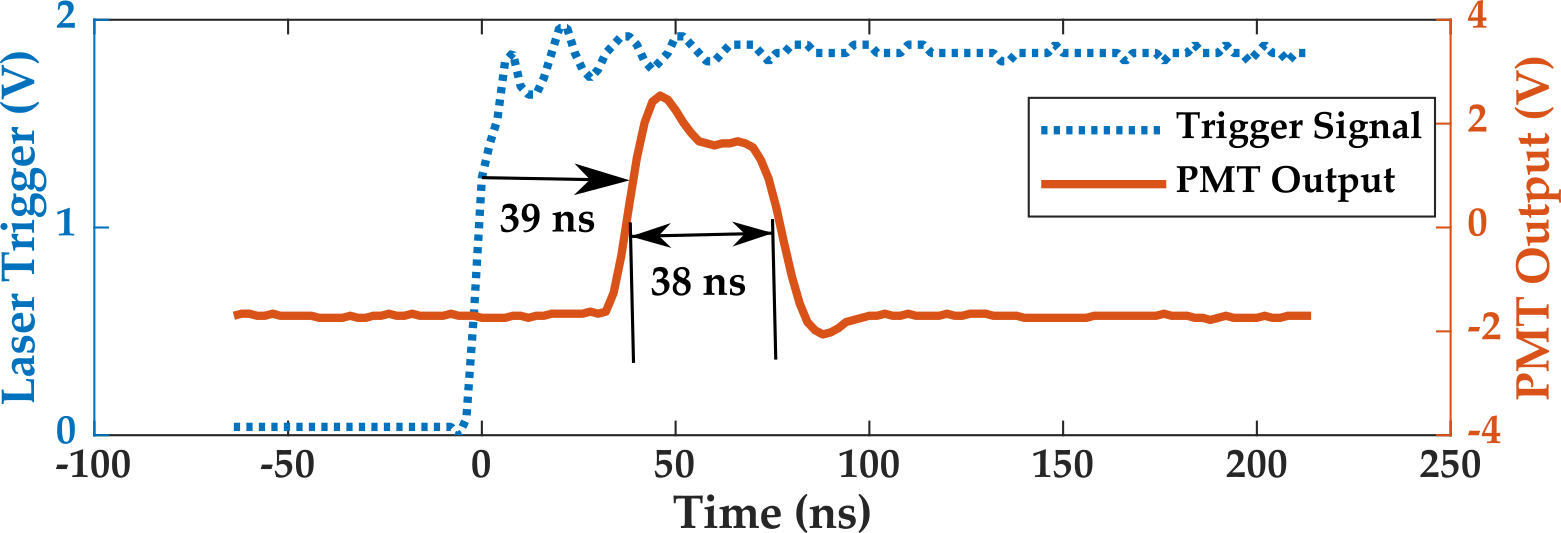}
    \vspace{-15pt}
    \caption{Photomultiplier tube (PMT) response to the laser pulse with a $38~ns$ pulse width setting showing a response delay of $39~ns$.}
    \vspace{-15pt}
    \label{fig:laser_delay}
\end{figure}

Although the internal delay of the laser $35\pm5~ns$ was provided by the manufacturer, the aggregated delay from the cables needed to be experimentally estimated. To that end, we modified the setup shown in Fig.~\ref{fig:dtof_model} by replacing the pg-SPAD with a commercial photomultiplier (PMT) detector and used an oscilloscope to estimate the system delay (see Fig.~\ref{fig:laser_delay}).

\subsection{Circuit-level Uncertainty, $\sigma_{TDC}$}
\begin{figure}
    \centering
    \includegraphics[width=\linewidth]{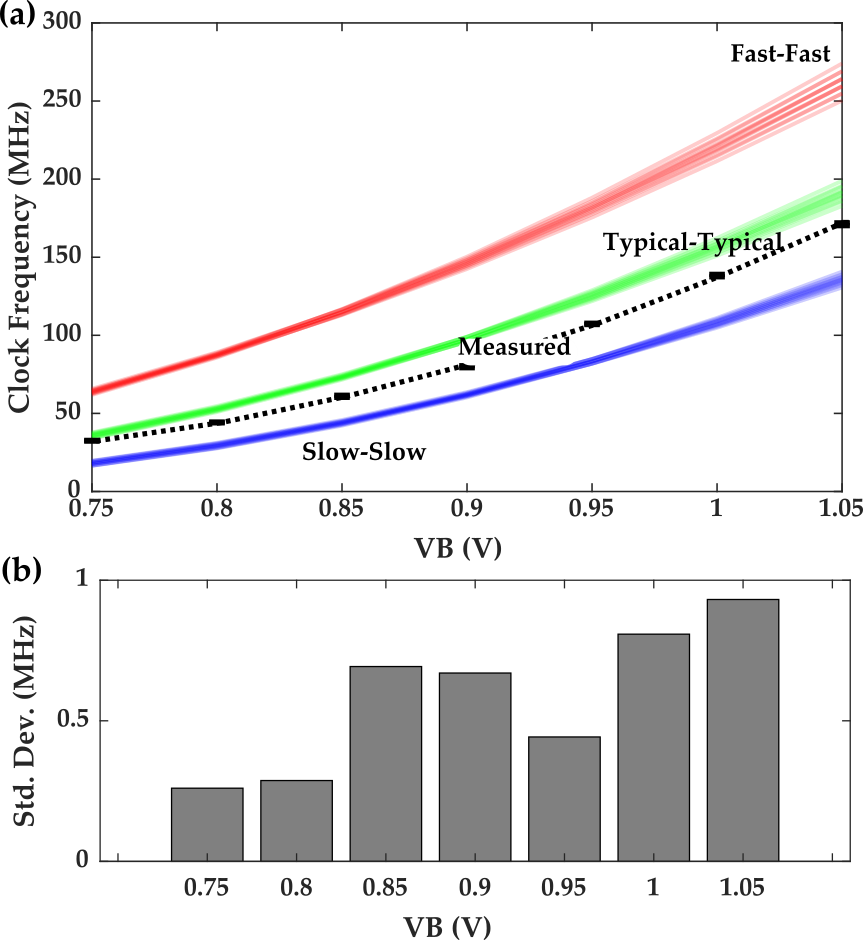}
    \vspace{-15pt}
    \caption{(a) Post-layout simulation of the TDC clock frequency under different corners (VDD = 3.25V (slow), 3.3V (typical), and 3.35V (fast); Temperature = $30^\circ\mathrm{C}, 40^\circ\mathrm{C}, 50^\circ\mathrm{C},$ and $60^\circ\mathrm{C}$), and (b) standard deviation of the measured clock frequency as a function of the clock control voltage, \textbf{VB} at $25^oC$.}
    \vspace{-15pt}
    \label{fig:clk}

\end{figure}

The variation in the TDC clock frequency introduces circuit-level uncertainty for counter-based TDCs. Here, we performed a post-layout simulation at different temperatures ($30^oC\sim60^oC$) and supply corners ($3.25~V\sim3.35~V$) to observe the effect on the clock frequency as we tuned the clock control bias (\textbf{VB}) (see Fig.~\ref{fig:clk} (a)). We have also overlaid the measured TDC frequency showing a good closeness to the typical-typical corner at room temperature. The useful \textbf{VB} range was found to be $[0.75\sim1.05]~V$. Beyond this voltage range, \textit{i.e.} $> 1.05~V$ or $< 0.75~V$, the clock frequency is unstable as it is running too fast to charge up the capacitive load or not running at all because the NMOS devices are close to cut-off, respectively.

Although the clock frequency variance increases with the base frequency (see Fig.~\ref{fig:clk} (b)), to utilize the highest timing resolution achievable from the TDC, we operate the clock at $150~MHz$ by keeping \textbf{VB} at $1.05~V$. This results in a maximum clock jitter of $\simeq 90~ps$. For a 12-bit TDC, this translates into 
\begin{equation}
\label{eq:tdc}
    \sigma_{TDC} = \sqrt{4096}*90~ps \simeq 5.7~ns.
\end{equation}

\begin{figure*}[h]
    \centering
    \includegraphics[width=.95\linewidth]{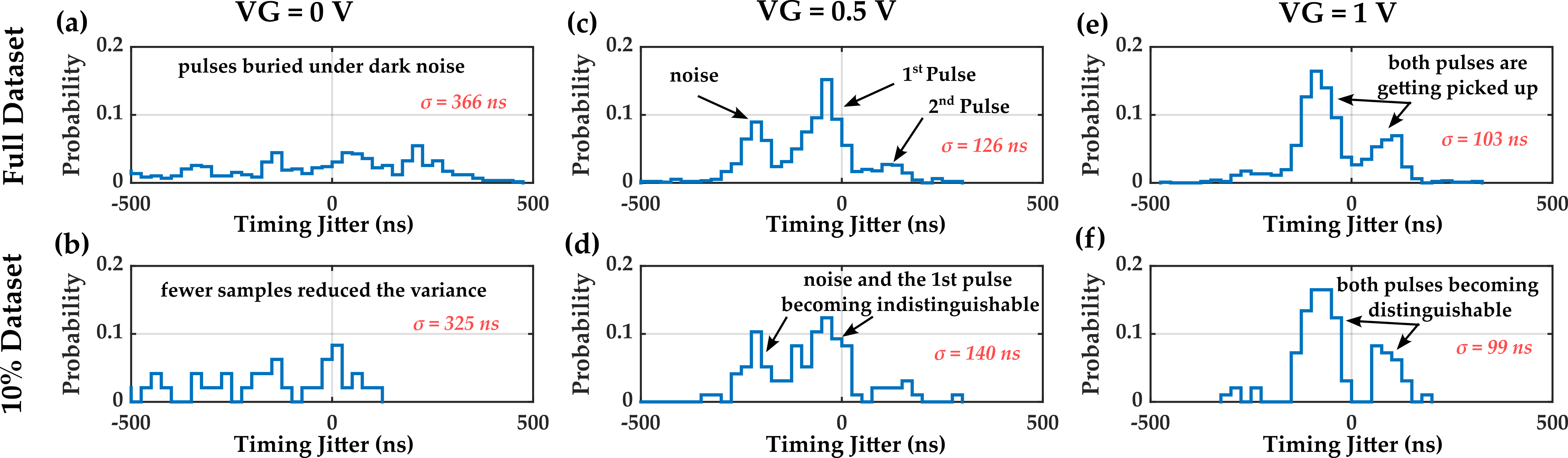}
    \vspace{-8pt}
    \caption{Histograms showing the avalanche events spread around the peak cluster, dominated by noise (VG = 0 V) and the photon pulses (VG = 0.5 and 1 V)}
    \vspace{-15pt}
    \label{fig:all_hist}
\end{figure*}

\subsection{Device-level Uncertainty, $\sigma_{SPAD}$}
Figure~\ref{fig:all_hist} shows normalized dToF histograms using full dataset (1000 repetitions shown on the top row) and $10\%$ of the dataset (first 100 repetitions shown on the bottom row) for gate voltages ranging from $0~V$ to $1~V$. Since perimeter gating trades sensitivity with noise, we limited it to $1~V$, which was enough to reduce the dark noise sufficiently for the given integration time~\cite{Sajal2022Perimeter-GatedProbability}. 

With $0~V$ on the gate, pg-SPAD behaves as a typical noisy SPAD. As a result, the histogram in Fig.~\ref{fig:all_hist} (a) shows a wide spread of dark events, making the mean of the histogram dominated by noise. The RMS standard deviation is found to be $366~ns$. With $10\%$ reduction of the data, the standard deviation reduced to $325~ns$. However, the pulses are still indistinguishable amidst the noise events as seen in Fig.~\ref{fig:all_hist} (b).

With increased gate voltage magnitude ($0~V\rightarrow0.5~V$), the dark count probability can be reduced so that the events corresponding to laser pulses are detected more frequently, leading to a standard deviation of $126~ns$. This is illustrated in Fig.~\ref{fig:all_hist} (c) for the full dataset. However, with the reduction of repetition (see Fig.~\ref{fig:all_hist} (d)), the dark events and the pulse events become indistinguishable which is also evident from the increased standard deviation of $140~ns$. This result has been anticipated through simulation as well in Fig.~\ref{fig:dtof_model} (b). It should be noted that we have used the maximum laser pulse width ($\sim38~ns$) and more than one pulse which are $200~ns$ apart to increase the photon detection probability since the fill-factor of the pg-SPADs was only $6\%$. Hence, we can see the detection of the $2^{nd}$ pulse when the dark noise was sufficiently suppressed.

Interestingly, further increase of the perimeter gate voltage ($0.5~V\rightarrow1~V$) shows higher suppression of the dark noise, resulting in a better detection of the laser pulses as shown in Fig.~\ref{fig:all_hist} (e). However, it is more interesting to note that even with $10\%$ of the data, the laser pulses are still resolved distinguishably (see Fig.~\ref{fig:all_hist} (f)). For both cases, the standard deviation is $103~ns$ and $99~ns$, respectively. Hence, pg-SPADs offer this unique benefit of dynamic noise suppression, allowing the reduction of pulse repetition for dToF measurements without compromising the accuracy, assuming other factors \textit{i.e.,} laser power, reflection, environmental conditions \textit{etc.} are held constant.

\begin{figure}
    \centering
    \includegraphics[width=.9\linewidth]{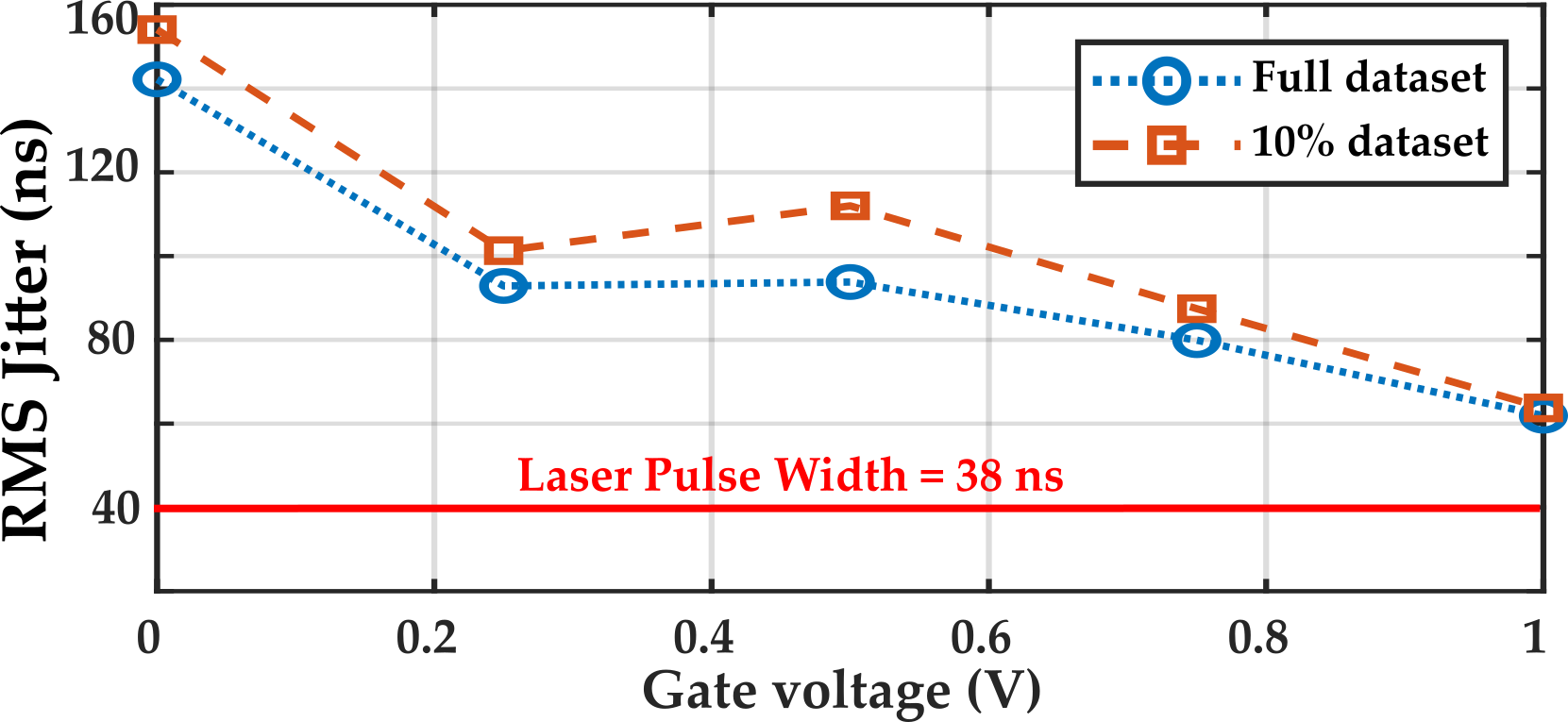}
    \vspace{-10pt}
    \caption{Effect of perimeter gate voltage on $\sigma_{total}$ in a simulated time-gated operation with a $200~ns$ gate window applied to the measured histogram.}
    \vspace{-8pt}
    \label{fig:perform}
\end{figure}

Although Fig.~\ref{fig:all_hist} visually presents the capabilities pg-SPAD's to reduce $\sigma_{total}$ in dToF measurements using multiple pulses, Fig.~\ref{fig:perform} specifically focused on the RMS jitter around the peak cluster corresponding to the $1^{st}$ pulse. In other words, Fig.~\ref{fig:perform} reports $\sigma_{total}$ \textit{if} a time gated operation of $200~ns$ active window were used with a $38~ns$ laser pulse. Time gated operation is one of the practical ways to reduce $\sigma_{total}$ by reducing the number of false detections caused by the dark carriers or the ambient photons. Hence, the RMS jitter reduced from $366~ns$ to $160~ns$ for VG = 0 V as seen in Fig.~\ref{fig:perform}. However, perimeter gating can still improve it further down to $\sim60~ns$ by reducing the dark noise within the time gated window as we see for VG = 1 V. By using Eqs.~\ref{eq:sigma}-\ref{eq:tdc}, we find
\begin{equation}
    \sigma_{SPAD} = \sqrt{60^2-5.7^2-.364^2} = 59.72~ns \simeq \sigma_{total}.
\end{equation}

\section{Limitations and Future Work}
Although we have shown that $\sigma_{total}$ is dominated by $\sigma_{SPAD}$ when $\sigma_{other}$ is neglected in practice, the laser pulse width \textit{i.e.}, $\sim38~ns$ sets the lower limit for it. Further improvement is achievable by using smaller pulse width, that would require increased number of repetition to compensate for the reduction in photon detection probability. This was difficult to achieve with the poor fill-factor ($6\%$) of the current prototype. Our future study aims to replace it with a higher fill-factor pg-SPAD imager. 

Another limitation is set by the TDC clock resolution and jitter. Using a delay line based TDC like Vernier TDCs, higher resolution can be achieved. However, in-pixel TDCs will directly trade-off with the fill factor of the pixel.

One explanation for the source of higher $\sigma_{SPAD}$ than the laser pulse limit is the afterpulsing. Since we have used one pg-SPAD due to readout limitation, we could not utilize co-incidence detection with the current setup. This is also left as a future study which will further reduce $\sigma_{SPAD}$ in addition to the benefit obtained through perimeter gating.

\section{Conclusions}
In this work, we have studied the effect of the perimeter gate voltage on dToF measurement accuracy. Although several works have previously predicted the benefits using simulations, we have experimentally demonstrated it by using a perimeter-gated SPAD (pg-SPAD) array fabricated in a $0.35~\mu m$ standard CMOS process. In particular, we have shown that the dToF measurement uncertainty ($\sigma_{total}$) in first photon detection based system can be substantially reduced by using pg-SPADs instead of the native devices. Perimeter gating can even be useful in time-gated operation to further improve measurement accuracy. Despite several system-level and readout limitations, our primary results showed promising outcome to further the study with high fill-factor pg-SPADs taking advantage of coincidence detection.


\newpage
\bibliographystyle{IEEEtranDOI}
\bibliography{main}

@inproceedings{4292664,
  author={Gudmundsson, Sigurjon Ami and Aanaes, Henrik and Larsen, Rasmus},
  booktitle={2007 International Symposium on Signals, Circuits and Systems}, 
  title={{Environmental Effects on Measurement Uncertainties of Time-of-Flight Cameras}}, 
  year={2007},
  volume={1},
  number={},
  pages={1-4},
  doi={10.1109/ISSCS.2007.4292664}}

@INPROCEEDINGS{6658237,
  author={Alahdab, Salim and Mäntyniemi, Antti and Kostamovaara, Juha},
  booktitle={2013 IEEE NoMe TDC}, 
  title={{Review of a Time-to-Digital Converter (TDC) based on Cyclic Time Domain Successive Approximation Interpolator Method with Sub-ps-Level Resolution}}, 
  year={2013},
  volume={},
  number={},
  pages={1-5},
  doi={10.1109/NoMeTDC.2013.6658237}}

@INPROCEEDINGS{6658231,
  author={Tamborini, D. and Markovic, B. and Tisa, S. and Villa, F. A. and Tosi, A.},
  booktitle={2013 NoMe TDC}, 
  title={{TDC with 1.5$\%$ DNL based on a Single-Stage Vernier Delay-Loop Fine Interpolation}}, 
  year={2013},
  volume={},
  number={},
  pages={1-6},
  doi={10.1109/NoMeTDC.2013.6658231}}

@INPROCEEDINGS{6737636,
  author={Abdelmejeed, M. and Guindi, R. and Abdel-Moneum, M.},
  booktitle={2013 4th Annual International Conference on Energy Aware Computing Systems and Applications (ICEAC)}, 
  title={{A Novel 10-Bit High-Throughput Two-Stage TDC with Reduced Power and Improved Linearity}}, 
  year={2013},
  volume={},
  number={},
  pages={50-54},
  doi={10.1109/ICEAC.2013.6737636}}

@INPROCEEDINGS{cnttdc3,
  author={Nakagawa, Shuya and Horikoshi, Kaito and Ishikuro, Hiroki},
  booktitle={2018 IEEE 61st International Midwest Symposium on Circuits and Systems (MWSCAS)}, 
  title={{A High-Resolution Time-Based Resistance-to-Digital Converter with TDC and Counter}}, 
  year={2018},
  volume={},
  number={},
  pages={242-245},
  doi={10.1109/MWSCAS.2018.8623991}}

@INPROCEEDINGS{cnttdc2,
  author={Choi, Hyoung-Taek and Kim, Young-Hwa and Kim, KwangSeok and Kim, Jaewook and Cho, SeongHwan},
  booktitle={2011 IEEE International Symposium on Radio-Frequency Integration Technology}, 
  title={{Time-Interleaved Single-Slope ADC using Counter-based Time-to-Digital Converter}}, 
  year={2011},
  volume={},
  number={},
  pages={185-188},
  doi={10.1109/RFIT.2011.6141768}}

@INPROCEEDINGS{cnttdc1,
  author={Yang, Jie and Zhao, Ye and Zhao, Jie},
  booktitle={2021 IEEE 3rd International Conference on Circuits and Systems (ICCS)}, 
  title={{A Digital Uncertainty-Tolerant Mismatch Compensation Method in a 17-channel Counter-Sampling Based TDC Architecture}}, 
  year={2021},
  volume={},
  number={},
  pages={180-184},
  doi={10.1109/ICCS52645.2021.9697126}}

@INPROCEEDINGS{10115088,
  author={Irfan, Nahin and Hasan, Sajid and Ara Shawkat, Shamim},
  booktitle={SoutheastCon 2023}, 
  title={{Modeling of Perimeter Gated SPAD-based Direct Time-of-Flight Sensor for Low Light LiDAR Applications}}, 
  year={2023},
  volume={},
  number={},
  pages={749-754},
  doi={10.1109/SoutheastCon51012.2023.10115088}}

@INPROCEEDINGS{10405999,
  author={Noyan, Utku and Lu, Sheung and Al-Shabili, Abdullah and Dandin, Marc and Chan, Stanley H. and Abshire, Pamela},
  booktitle={2023 IEEE 66th International Midwest Symposium on Circuits and Systems (MWSCAS)}, 
  title={{Pulsed ToF LiDAR-Based Depth Imaging: SPAD Circuit Considerations and Simulation Study}}, 
  year={2023},
  volume={},
  number={},
  pages={875-879},
  doi={10.1109/MWSCAS57524.2023.10405999}}

@article{Dandin2010,
    title = {{Characterization of Single-Photon Avalanche Diodes in a 0.5 {$\mu$}m Standard CMOS Process—Part 1: Perimeter Breakdown Suppression}},
    year = {2010},
    journal = {IEEE Sensors Journal},
    author = {Dandin, Marc and Akturk, Akin and Nouri, Babak and Goldsman, Neil and Abshire, Pamela},
    number = {11},
    pages = {1682--1690},
    volume = {10},
    doi = {10.1109/JSEN.2010.2046163},
    issn = {1530-437X}
}

@article{Dandin2016,
    title = {{Characterization of Single-Photon Avalanche Diodes in a 0.5 {$\mu$}m Standard CMOS Process—Part 2: Equivalent Circuit Model and Geiger Mode Readout}},
    year = {2016},
    journal = {IEEE Sensors Journal},
    author = {Dandin, Marc and Habib, Mohammad Habib Ullah and Nouri, Babak and Abshire, Pamela and McFarlane, Nicole},
    number = {9},
    month = {5},
    pages = {3075--3083},
    volume = {16},
    doi = {10.1109/JSEN.2016.2526665},
    issn = {1530-437X}
}

@article{Dandin2012a,
    title = {{High Signal-to-Noise Ratio Avalanche Photodiodes with Perimeter Field Gate and Active Readout}},
    year = {2012},
    journal = {IEEE Electron Device Letters},
    author = {Dandin, Marc and Abshire, Pamela},
    number = {4},
    pages = {570--572},
    volume = {33},
    doi = {10.1109/LED.2012.2186112},
    issn = {0741-3106}
}

@inproceedings{Nouri2012,
    title = {{Large-Area Low-Noise Single-Photon Avalanche Diodes in Standard CMOS}},
    year = {2012},
    booktitle = {2012 IEEE Sensors},
    author = {Nouri, Babak and Dandin, Marc and Abshire, Pamela},
    pages = {1--5},
    publisher = {IEEE},
    isbn = {978-1-4577-1767-3},
    doi = {10.1109/ICSENS.2012.6411365}
}

@article{Richardson2011ScaleableTechnology,
    title = {{Scaleable Single-Photon Avalanche Diode Structures in Nanometer CMOS Technology}},
    year = {2011},
    journal = {IEEE Transactions on Electron Devices},
    author = {Richardson, Justin A. and Webster, Eric A. G. and Grant, Lindsay A. and Henderson, Robert K.},
    number = {7},
    pages = {2028--2035},
    volume = {58},
    doi = {10.1109/TED.2011.2141138},
    issn = {0018-9383}
}

@article{Finkelstein2006STI-BoundedTechnology,
    title = {{STI-Bounded Single-Photon Avalanche Diode in a Deep-Submicrometer CMOS Technology}},
    year = {2006},
    journal = {IEEE Electron Device Letters},
    author = {Finkelstein, H. and Hsu, M.J. and Esener, S.C.},
    number = {11},
    pages = {887--889},
    volume = {27},
    doi = {10.1109/LED.2006.883560},
    issn = {0741-3106}
}

@article{Palubiak2011High-SpeedApplications,
    title = {{High-Speed, Single-Photon Avalanche-Photodiode Imager for Biomedical Applications}},
    year = {2011},
    journal = {IEEE Sensors Journal},
    author = {Palubiak, Darek and El-Desouki, Munir M. and Marinov, Ognian and Deen, M. Jamal and Fang, Qiyin},
    number = {10},
    pages = {2401--2412},
    volume = {11},
    doi = {10.1109/JSEN.2011.2123090},
    issn = {1530-437X}
}

@INPROCEEDINGS{dtofintro3,
  author={Ximenes, Augusto Ronchini and Padmanabhan, Preethi and Lee, Myung-Jae and Yamashita, Yuichiro and Yaung, D. N. and Charbon, Edoardo},
  booktitle={2018 IEEE International Solid-State Circuits Conference - (ISSCC)}, 
  title={{A 256×256 45/65nm 3D-Stacked SPAD-based Direct TOF Image Sensor for LiDAR Applications with Optical Polar Modulation for up to 18.6dB Interference Suppression}}, 
  year={2018},
  volume={},
  number={},
  pages={96-98},
  doi={10.1109/ISSCC.2018.8310201}}

@INPROCEEDINGS{dtofintro2,
  author={Faizan, Sohail and Bisen, Minal and Jainwal, Kapil},
  booktitle={2023 IEEE International Symposium on Circuits and Systems (ISCAS)}, 
  title={{A SPAD Based dToF Pixel with Log/Linear Multi-mode Operation for LiDAR Applications}}, 
  year={2023},
  volume={},
  number={},
  pages={1-5},
  doi={10.1109/ISCAS46773.2023.10181536}}

@INPROCEEDINGS{dtofintro1,
  author={Kim, Eo-Jin and Park, Hyo-Sung and Choi, Woo-Young and Lee, Myung-Jae},
  booktitle={2024 IEEE International Conference on Consumer Electronics-Asia (ICCE-Asia)}, 
  title={{Direct Time-of-Flight Sensor System Based on SPAD IC}}, 
  year={2024},
  volume={},
  number={},
  pages={1-4},
  doi={10.1109/ICCE-Asia63397.2024.10773705}}

@Article{first_photon,
AUTHOR = {Albert, Konstantin and Ligges, Manuel and Henschke, Andre and Ruskowski, Jennifer and De Zoysa, Menaka and Noda, Susumu and Grabmaier, Anton},
TITLE = {{Performance Comparison of Multipixel Biaxial Scanning Direct Time-of-Flight Light Detection and Ranging Systems With and Without Imaging Optics}},
JOURNAL = {Sensors},
VOLUME = {25},
YEAR = {2025},
NUMBER = {10},
ARTICLE-NUMBER = {3229},
PubMedID = {40432021},
ISSN = {1424-8220},
DOI = {10.3390/s25103229}
}

@Article{dtof_model,
AUTHOR = {Padmanabhan, Preethi and Zhang, Chao and Charbon, Edoardo},
TITLE = {{Modeling and Analysis of a Direct Time-of-Flight Sensor Architecture for LiDAR Applications}},
JOURNAL = {Sensors},
VOLUME = {19},
YEAR = {2019},
NUMBER = {24},
ARTICLE-NUMBER = {5464},
PubMedID = {31835807},
ISSN = {1424-8220},
DOI = {10.3390/s19245464}
}

@ARTICLE{dtofaccuracy1,
  author={Koerner, Lucas J.},
  journal={IEEE Transactions on Instrumentation and Measurement}, 
  title={{Models of Direct Time-of-Flight Sensor Precision That Enable Optimal Design and Dynamic Configuration}}, 
  year={2021},
  volume={70},
  number={},
  pages={1-9},
  doi={10.1109/TIM.2021.3073684}}

@INPROCEEDINGS{tdc4,
  author={Qi, Xinren and Wang, Yonggang},
  booktitle={2024 IEEE International Instrumentation and Measurement Technology Conference (I2MTC)}, 
  title={{Design of A High Performance Time-to-Digital Converter with Zero Dead Time on Xilinx FPGA}}, 
  year={2024},
  volume={},
  number={},
  pages={1-6},
  doi={10.1109/I2MTC60896.2024.10560688}}

@ARTICLE{tdc3,
  author={Qin, Jiajun and Zhao, Junbo and Li, Zhuang and Li, Jiaming and Guo, Donglei and Zhao, Lei},
  journal={IEEE Transactions on Nuclear Science}, 
  title={{A Sub-10-ps Resolution Multichannel Time-to-Digital Converter With Two-Stage Interpolation for High-Energy Physics Applications}}, 
  year={2025},
  volume={72},
  number={3},
  pages={325-332},
  doi={10.1109/TNS.2024.3473869}}

@INPROCEEDINGS{tdc2,
  author={Fang, Shilong and Cai, Yanlong and Huang, Yingqi and Yan, Chenggang and Liu, Weiqiang},
  booktitle={2024 9th International Conference on Integrated Circuits and Microsystems (ICICM)}, 
  title={{A 350MHz, 10bit, 500-fs Resolution Asynchronous Pipelined-SAR Time-to-Digital Converter}}, 
  year={2024},
  volume={},
  number={},
  pages={429-433},
  doi={10.1109/ICICM63644.2024.10814216}}

@ARTICLE{tdc1,
  author={Liu, Ruqing and Li, Feng and Zhu, Jingguo and Jiang, Yan and Jiang, Chenghao and Hu, Tao},
  journal={Chinese Journal of Electronics}, 
  title={{A High-Resolution Calibration Method for Time-to-Digital Converter of Lidar}}, 
  year={2025},
  volume={34},
  number={1},
  pages={222-228},
  doi={10.23919/cje.2023.00.237}}

@ARTICLE{dtof4,
  author={Dabidian, Sobhan and Tafaghodi Jami, Sadra and Kavehvash, Zahra and Fotowat-Ahmady, Ali},
  journal={IEEE Sensors Journal}, 
  title={{Direct Time-of-Flight (d-ToF) Pulsed LiDAR Sensor With Simultaneous Noise and Interference Suppression}}, 
  year={2024},
  volume={24},
  number={17},
  pages={27578-27586},
  doi={10.1109/JSEN.2024.3425173}}

@ARTICLE{dtof3,
  author={Gyongy, Istvan and Erdogan, Ahmet T. and Dutton, Neale A.W. and Martín, Germán Mora and Gorman, Alistair and Mai, Hanning and Rocca, Francesco Mattioli Della and Henderson, Robert K.},
  journal={IEEE Journal of Selected Topics in Quantum Electronics}, 
  title={{A Direct Time-of-Flight Image Sensor With In-Pixel Surface Detection and Dynamic Vision}}, 
  year={2024},
  volume={30},
  number={1: Single-Photon Technologies and Applications},
  pages={1-11},
  doi={10.1109/JSTQE.2023.3238520}}

@ARTICLE{spadpuf,
  author={Sajal, Md Sakibur and Dandin, Marc},
  journal={IEEE Transactions on Circuits and Systems I: Regular Papers}, 
  title={{PG-SPAD PUFs: Reconfigurable Physically Unclonable Functions Using Perimeter-Gated Single-Photon Avalanche Diodes}}, 
  year={2025},
  volume={},
  number={},
  pages={1-12},
  doi={10.1109/TCSI.2025.3607762}}

@ARTICLE{dtof2,
  author={Niclass, Cristiano and Soga, Mineki and Matsubara, Hiroyuki and Ogawa, Masaru and Kagami, Manabu},
  journal={IEEE Journal of Solid-State Circuits}, 
  title={{A 0.18-$\mu$m CMOS SoC for a 100-m-Range 10-Frame/s 200 $\,\times\,$96-Pixel Time-of-Flight Depth Sensor}}, 
  year={2014},
  volume={49},
  number={1},
  pages={315-330},
  doi={10.1109/JSSC.2013.2284352}}

@INPROCEEDINGS{dtof1,
  author={Al Abbas, T. and Dutton, N. A. W. and Almer, O. and Pellegrini, S. and Henrion, Y. and Henderson, R. K.},
  booktitle={2016 IEEE International Electron Devices Meeting (IEDM)}, 
  title={{Backside Illuminated SPAD Image Sensor with 7.83 $\mu$m Pitch in 3D-stacked CMOS Technology}}, 
  year={2016},
  volume={},
  number={},
  pages={8.1.1-8.1.4},
  doi={10.1109/IEDM.2016.7838372}}

@inproceedings{Sajal2022Perimeter-GatedProbability,
    title = {{Perimeter-Gated Single-Photon Avalanche Diode Imager with Vanishing Room Temperature Dark Count Probability}},
    year = {2022},
    booktitle = {2022 29th IEEE International Conference on Electronics, Circuits and Systems (ICECS)},
    author = {Sajal, Md. Sakibur and Lin, Kai-Chun and Senevirathna, Bathiya and Lu, Sheung and Dandin, Marc},
    pages = {1--4},
    publisher = {IEEE},
    doi = {10.1109/ICECS202256217.2022.9970824}
}

@article{Sajal2024TrueDiode,
    title = {{True Random Number Generation Using Dark Noise Modulation of a Single-Photon Avalanche Diode}},
    year = {2024},
    journal = {IEEE Transactions on Circuits and Systems II: Express Briefs},
    author = {Sajal, Md. Sakibur and Dandin, Marc},
    number = {3},
    month = {3},
    pages = {1586--1590},
    volume = {71},
    doi = {10.1109/TCSII.2023.3347735},
    issn = {1549-7747}
}

\end{document}